\documentclass{article}
\usepackage{ijcai24}

\usepackage{times}
\usepackage{soul}
\usepackage{url}
\usepackage[hidelinks]{hyperref}
\usepackage[utf8]{inputenc}
\usepackage[small]{caption}
\usepackage{graphicx}
\usepackage{amsmath}
\usepackage{amsthm}
\usepackage{booktabs}
\usepackage{algorithm}
\usepackage{algorithmic}
\usepackage[switch]{lineno}

\usepackage{fancyhdr}

\fancypagestyle{plain}{
  \fancyhf{}
  \fancyfoot[C]{\thepage}

}

\usepackage{amsmath}
\usepackage[dvipsnames]{xcolor}
\usepackage{xspace}
\usepackage{tabularx}

\usepackage{tcolorbox}

\newcommand{\Gitassist}{Git-Assistant\xspace}
\newcommand{\Gitpddlassist}{Hybrid-planner\xspace}

\title{Git-Assistant: Planning-Based Support for Updating Git Repositories}

\author{
Alfredo Garrachón Ruiz
\and
Tomás de la Rosa\and
Daniel Borrajo\\
\affiliations
AI Research, JPMorganChase\\
}

\begin{document}

\maketitle

\thispagestyle{plain}

\begin{abstract}
Version control systems are essential for collaborative software development, yet tools like git remain challenging for many practitioners. Recent advances in Large Language Models (LLMs) offer promising capabilities for interpreting developer intent, but their effectiveness in repository management tasks is limited by the need for formal reasoning. This work introduces \Gitassist, an AI-based assistant that combines LLMs with automated planning to support developers in executing non-trivial git operations. The assistant analyzes repository context, translates natural language requests into actionable command sequences, and incorporates planning techniques to ensure correctness and safety. We present a systematic evaluation methodology using synthetic and randomized git environments, comparing the performance of LLM-only and planning-augmented variants across multiple metrics. Experimental results demonstrate that integrating formal reasoning with LLMs improves reliability and reduces errors in repository management, highlighting the potential of hybrid AI approaches for intelligent developer assistance.
\end{abstract}

\section{Introduction}
Version control systems are fundamental to modern software development, with git standing out as the most widely adopted tool for managing collaborative workflows, code integration, and project history. However, despite its central role, git is often perceived as difficult to master. Both new and experienced developers encounter significant challenges, especially when dealing with complex operations such as branching, merging, and resolving conflicts~\cite{DeRosso_2013_GitConceptual}.
Empirical studies and analyses of developer activity on platforms like GitHub and Stack Overflow~\cite{yang2022developers,milliken2021behavioral} consistently show that misunderstandings and mistakes are common in everyday git usage. Developers frequently make errors in repository management and pull request workflows, highlighting the cognitive demands and error-prone nature of manual git operations~\cite{kinsman2021software}. These difficulties are further compounded by git’s complex command-line interface and distributed model, which require users to memorize sequences of commands and understand subtle repository states~\cite{deRosso_2016_GitMisfits}.  Modern git clients and IDE plugins offer some support through the git graph (i.e., visual representations of repository history) and simplified menus of git commands, but they often fall short when it comes to constructing the full sequence of commands needed to achieve a specific repository state.

Recent advances in Large Language Models (LLMs) have demonstrated that these models are highly effective at understanding and generating natural language \cite{minaee2024_LLMsurvey}. This makes LLMs a promising foundation for building AI-based assistants that can interpret developer intent and provide contextual guidance for git operations. By bridging the gap between natural language requests and precise git command sequences, LLMs have the potential to lower the barrier to effective git usage and reduce the likelihood of mistakes. 

Nevertheless, while LLMs are good at language understanding, the complexity of repository management tasks suggests that additional formal reasoning techniques, such as automated planning, may be needed to ensure the correctness and safety of generated command sequences. There is a growing need for intelligent, context-aware assistants that can analyze the current repository state, recommend operations, and help prevent users from reaching undesirable or inconsistent git states.

In this work, we present \Gitassist, a CLI assistant that receives developers' git-related requests and provides a sequence of git commands to address those requests. To understand the benefits that \Gitassist can provide, we address the following research questions: (1) Does incorporating repository context improve the accuracy of LLMs answers to natural language requests for git command preparation? (2) Can formal reasoning methods, such as automated planning, enhance the reliability and safety of generated git command sequences? (3) To what extent does reasoning-based assistance help prevent users from reaching problematic or unintended git states?


In the following sections, we discuss related work, describe the design of various assistant versions, and present the methodological procedures used to evaluate them, along with the results of our evaluation.



\section{Related Work}
A git workflow is a structured set of guidelines that governs how developers collaborate using git. Popular workflows, such as Gitflow~\cite{gitflow} and Trunk-based development, are widely adopted by development teams, and several analyses compare these workflows based on their features~\cite{rios2022unifying}. Although workflows establish a team agreement that restricts the universe of applicable commands, the developer is still responsible for devising the sequence of these commands. 

Current automation efforts mainly enhance continuous integration and deployment (CI/CD) processes~\cite{wessel2023github} on collaborative platforms, such as triggering pipelines after commits. These solutions focus on software delivery rather than automating complex version control management tasks. Some methods train models on Stack Overflow posts to suggest git commands~\cite{jia2023_gitrecommend} or propose similar scenarios~\cite{shen2024richen} by matching user queries in natural language with existing examples. While these approaches provide relevant commands and examples, they do not generate custom command sequences tailored to the specific repository the developer is working on.

A notable prior approach~\cite{bachmann2021_gitmodel} modeled the complete git commit graph using the Planning Domain Definition Language (PDDL)~\cite{PDDL}, where actions are defined as graph mutations and the objective is to reach a specified target graph state. While this formulation enables planning for repository management, it assumes that developer intent can be precisely captured as a target graph. However, this does not align with the more flexible and abstract goals typically expressed by practitioners. Furthermore, this work primarily addresses local graph transformations and does not account for operations involving the synchronization of remote repositories, such as pull or push commands. In contrast, our work focuses on representing only the essential repository information required to support a broad set of git commands and mapping natural language requests to goals, thereby enhancing practical applicability.

\section{Background}

\begin{figure}
    \centering
    \includegraphics[width=0.9\linewidth]{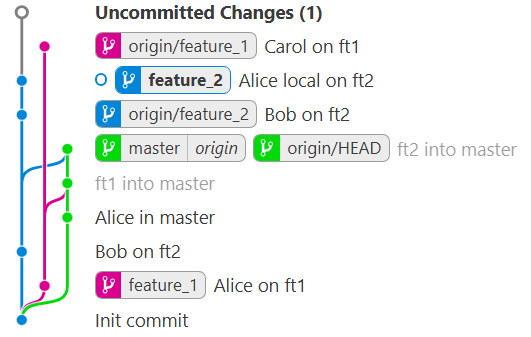}
    \begin{tcolorbox}
    
    \begin{footnotesize}
    \textbf{Intent}: add my modified files to the latest version of feature\_1 and push them to the repo
    \vspace{0.1cm} 
    
    \textbf{Git commands}: 
    
    1. git fetch origin
    
    2. git stash
        
    3. git checkout feature\_1
    
    4. git pull origin feature\_1
  
    5. git stash pop \hspace*{2em} (\textit{Alice intervention in case of conflicts)}
  
    6. git add -u
  
    7. git commit -m "Committing changes"
  
    8. git push origin feature\_1
  
    \end{footnotesize}
        
    \end{tcolorbox}
    \caption{(Top) Git graph example shown from local (\textit{Alice}) perspective. 
    \textcolor{NavyBlue}{feature\_2} is \textit{ahead} of origin, 
     \textcolor{magenta}{feature\_1} is \textit{behind} and \textcolor{ForestGreen}{master} is \textit{in sync}.
     (Bottom) Commands that Alice applies to achieve her goal.
    }
    \label{fig:git_graph}
\end{figure}

Git organizes project files and their history within a repository, where changes are recorded as \textit{commits}. Developers work on \textit{branches} to isolate features or fixes, and synchronize their work through a remote server, typically referred to as \textit{origin}. 
The local environment consists of the \textit{working tree} (i.e., the directory of editable files) and the index (or \textit{staging area}), which holds changes ready to be committed. Files are categorized as tracked, untracked, modified or deleted, which determine the behavior of some command parameters. Core operations such as \textit{commit}, \textit{pull}, \textit{push}, and \textit{merge} are essential for managing changes and collaborating with others. The \textit{rebase} operation allows developers to move or combine a sequence of commits, re-writing the history of existing branches. The \textit{reset} command enables rolling back the repository to a previous commit, either discarding or preserving changes in the working tree. A developer can also \textit{squash} multiple commits into a single one, helping to simplify the commit history.

Git updates files at a granular level, comparing them line by line to identify differences. When changes from different sources overlap, merge conflicts are detected. These conflicts may require manual intervention, but git also supports automated resolution strategies, such as preferring local changes or already committed versions.
Features like \textit{stash} (temporal storage of local changes) and \textit{tag} (marking a significant point in the project's history) offer additional management capabilities. 

Internally, git represents repository history as a directed acyclic graph, such as in Figure~\ref{fig:git_graph} (top).
Both local and origin branches are pointers to commits, but local branches exist on the user's machine while origin branches represent the state of branches on the remote repository. Origin branches are updated when the user fetches from the remote, allowing local branches to track changes made by others.
Local branches can be pushed to or pulled from origin, enabling collaboration and synchronization between individual contributors and the shared repository. As a result, local branches may be ahead of, behind, synchronized with, or diverged from their origin branch, which impacts workflow and requires careful management.

Git’s rich set of commands and parameters, along with its internal checks and preconditions, can make certain operations complex, particularly for users with limited experience. Consider, for example, the scenario depicted in Figure~\ref{fig:git_graph} (top), and suppose that uncommitted changes have modified files that are shared across multiple branches. A seemingly simple request such as \textit{"add my modified files to the latest version of feature\_1 and push them to the repository"} will require a sequence of eight commands, as illustrated in Figure~\ref{fig:git_graph} (bottom). Other basic functionalities, such as squashing two commits or rebasing a branch, involve rewriting the history in origin and necessarily require pushing with the \textit{--force} option. For inexperienced users, this can lead to the intimidating feeling of potentially breaking the repository.

\section{Architecture}
\label{architecture}
\begin{figure}[hbt]
    \centering
    \includegraphics[width=0.8\linewidth]{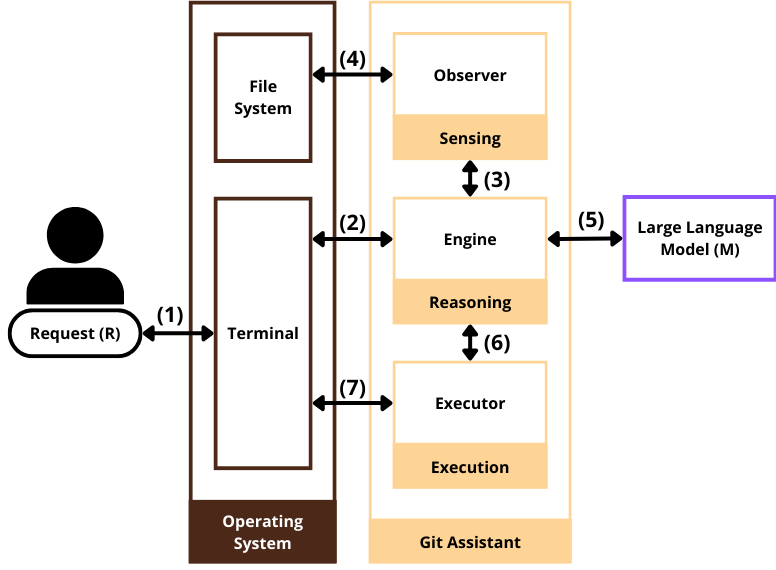}
    \caption{Architecture of the Git-Assistant agent}
    \label{fig:architecture}
\end{figure}

\Gitassist can be abstracted as the composition of three modules, each fulfilling a necessary function to operate as an agent (see Figure~\ref{fig:architecture} for representation and flow orchestration). The \textbf{Observer} performs the \textit{sensing} of the environment, the \textbf{Engine} carries out the \textit{reasoning} for achieving goals as well as the interaction with the user. And  the \textbf{Executor} is responsible for the \textit{execution} of the plan (git commands).

When a user requests to solve a git task, the \textit{Engine} first identifies whether the user's input is a request in natural language or a git command. If it is a git command, the module executes it and returns the information to the user for transparent interaction. If it is a task in natural language, it first gets the context provided by the \textit{Observer}, which extracts all the relevant information from the target git repository. This includes  the git graph, branch states and working tree thereby obtaining a comprehensive representation of the repository. Then, the \textit{Engine} reasons about the user intent and the available context,  to obtain the set of commands necessary to satisfy the request. This module has two variants: (1) using a standalone LLM, or (2) using a hybrid approach, an LLM together with an automated planner used as a solver.

Once the commands to be executed are obtained, they are passed to the \textit{Executor}, which confirms with the user the proposed commands. On confirmation, it executes each command within the operating system, changing the state of local and remote repositories. 

Next, we delve into each of the components. 

\subsection{Observer}
The decision regarding which sequence 
of commands is appropriate for a user request largely depends on the current state of the local repository. Therefore, the Observer uses native git commands and the \textit{gitpython} library to extract the information that allows us to build the context of the current state. The state $S$ for a repository includes:
\begin{itemize}
    \item \textbf{Local status}: the current branch and the list of modified or untracked files.
    \item \textbf{Branch locations}: for each branch, its presence in local or remote repositories.
    \item \textbf{Branch relations}: the synchronization status of local branches with their remote counterparts (ahead, behind, synchronized, or diverged).
    \item \textbf{Last commit info}: the hash identifiers and authors of the latest commit for each branch.
    \item \textbf{Git graph}: the text representation of the repository graph provided by the git log command.
\end{itemize}

\subsection{LLM-based Engine}
\begin{figure*}[hbt]
    \centering
    \includegraphics[width=1\linewidth]{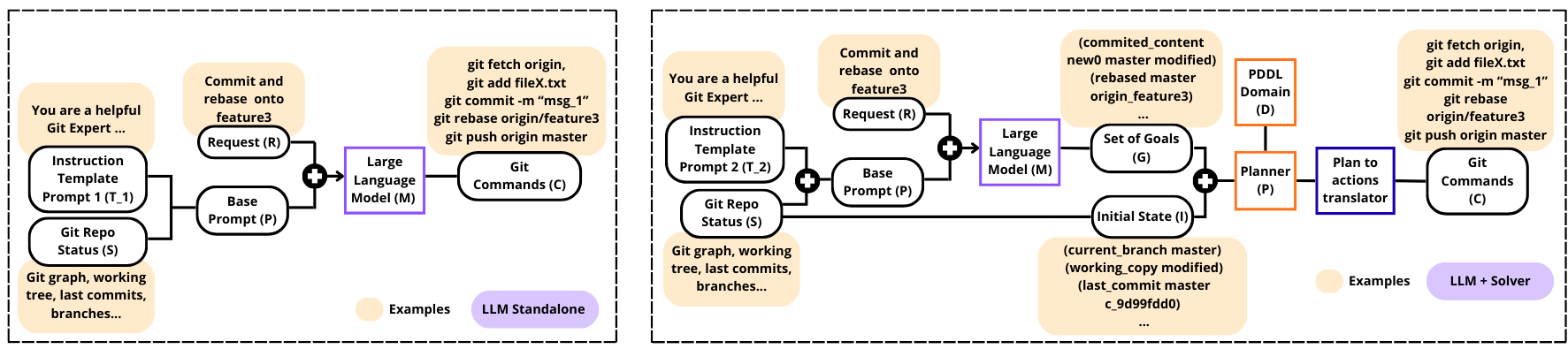}
    \caption{Diagram of the two variants of the Engine module}
    \label{fig:reason_module}
\end{figure*}

In the first version of the Engine, referred to as \textit{LLM-based} (see Figure~\ref{fig:reason_module}, left), the request is processed exclusively by an LLM. To facilitate this, an Instruction Template Prompt ($T_1$) has been designed, outlining the task, specifying relevant constraints, and including few-shot examples. This $T_1$ is nested with the repository state $S$ from the \textit{Observer} to create a complete context to the LLM, thus generating the base prompt ($P$). The base prompt $P$, together with the user request $R$, is fed to the LLM, which outputs the set of git commands ($C$) to be executed. See Appendix~\ref{appendix:prompts} for prompt template details.

\subsection{Hybrid Planner}

The second variant is a planning-based assistant (see Figure~\ref{fig:reason_module} right), named \textit{Hybrid-planner}. In this approach, the user's request is formulated as an automated planning task with the help of an LLM. This approach guarantees that, if a solution exists, the planner will identify a valid sequence of operations. The planning task is defined by the tuple $(D,  I, G)$, where $D$ denotes the Git domain specified in PDDL, $I$ is the initial state, and $G$ represents the desired goals.

\begin{table*}[ht]
\centering
\small 
\begin{tabularx}{\textwidth}{|l|X|X|l|}
\hline
\textbf{Group} & \textbf{Actions} & \textbf{Purpose/Transition Summary} & \textbf{Phase(s)} \\
\hline\hline 
Phase Management & git\_fetch, move\_to\_operate, move\_to\_pushing & Transition between flow phases (start $\rightarrow$ updating $\rightarrow$ operating $\rightarrow$ pushing  &  \\
\hline
Branch Operations & create\_branch, checkout\_branch, checkout\_remote\_branch & Create or switch branches, including remote branches & any \\
\hline
Down Synchronization & git\_pull\_behind, git\_pull\_diverged & Update local branch from remote & updating \\
\hline
Content Staging & add\_modified\_content, add\_specific\_content, add\_untracked\_content & Add different classes of content to staging area & operating \\
\hline
Commit Operations & git\_commit, git\_commit\_content & Create commits with staged content & operating \\
\hline
Merge \& Rebase Operations & git\_merge, git\_merge\_local, git\_rebase & Merge branches, re-apply commits on top of branches & operating \\
\hline
History Editing & squash\_commits, git\_reset\_count & squash commits, reset branch history & operating \\
\hline
Tagging & git\_tag & Tag significant commits & operating \\
\hline
Up Synchronization & git\_push, git\_push\_new\_branch, git\_force\_push, git\_diverged\_push  & Push local changes to remote, create or force update remote & pushing \\
\hline
Stash Management & git\_stash, git\_pop\_stash & Save changes to stash (start), restore from stash (pushing) & start, pushing \\
\hline

\end{tabularx}
\caption{Functional grouping of PDDL actions for git-tasks domain, with associated phases.}
\label{tab:git-tasks-groups}
\end{table*}

The domain encodes actions that modify the repository's state space. These actions correspond both to individual git commands and to phase transition actions, which enable the execution of operations in distinct phases. Table~\ref{tab:git-tasks-groups} presents the list of actions, organized by high-level git functionalities.  See Appendix~\ref{appendix:git-domain} for additional details.

The initial state $I$ is a PDDL translation of the repository state $S$. Local status, branch locations and relations, as well as last commit information, have a straightforward representation in predicate logic, which is generated programmatically. Instead of explicitly encoding the git graph, we represent commits that can be safely squashed and/or reverted as additional facts in the initial state.

For  \Gitpddlassist, the LLM is responsible for interpreting the user’s request and producing the set of goals $G$. We argue that identifying goals, such as $\mathsf{(merged\ master\ feature\_1)}$, is much simpler than deriving the steps required to achieve them. On one hand, goals are directly related to what developers typically want; on the other hand, the LLM does not need to reason about preconditions, such as being on the target branch. To implement this, we developed an additional Instruction Template Prompt ($T_2$), which instructs the LLM to obtain the goals $G$ and a new objects mapping $M$. The prompt includes explanations and examples on how to interpret requests and the corresponding PDDL representations of the goals for high-level git functionalities. The mapping $M$ is used to identify elements that are not named at the time of creating the initial state. For instance, consider the request "create a branch feature-ui from my current branch", with master as the current branch. The branch to be created does not exist in $I$, but 'new\_branch' is a PDDL object intentionally included in it. Here, $G=[(\mathsf{created\_from~new\_branch~ master})]$. The mapping ($\mathsf{new\_branch} \rightarrow$ "feature-ui"), provided in the LLM output, is then used to fill in the branch creation parameter.

The pair $(I, G)$ is typically specified in a PDDL problem. Thus, the task $(D, I, G)$, comprising a fixed git domain and a per-request problem, is passed to a PDDL-compatible automated planner to obtain a plan containing the sequence of actions required to achieve the goals. This plan is deterministically translated into the corresponding sequence of git commands $C$, thereby providing the necessary commands to fulfill the user's request.

\subsection{Executor}
Lastly, this module is the one in charge of executing the git commands $C$ obtained by means of the Enginge using either its variants. This module incorporates two extra capabilities: (1) an "interactive capability" that detects whenever there is a need for the user input (like resolving conflicts), returning the control of the execution to the user to resolve those, and then continue with the rest of the execution. (2) the "auto-message commit capability" that detects whenever a commit appears during the commands execution and automatically generates a proper commit message based on the differences between the last commit and the current state.

\section{Material and Methods}
Evaluating the effectiveness of a git assistant presents several challenges. First, the diversity of git command sequences capable of achieving the same set of goals and the implicit side effects complicate the establishment of standardized evaluation criteria. Consider, for instance, that the running example in Figure~\ref{fig:git_graph} would have an alternative solution: committing contents to current feature\_2 and then merging into feature\_1. This intermediate commit (side effect)  depends on how the assistant interprets the user's request. On the other hand, conflict resolution within git workflows frequently requires human judgment, as decisions regarding the handling of merge conflicts or rebase issues are context-dependent and often subjective. To our knowledge, the literature lacks suitable benchmarks specifically designed for our task, further hindering objective comparison and progress in this domain. In response, we have developed a methodology to automate the continuous evaluation of such tasks, enabling systematic and reproducible assessment.

This methodology comprises the following sequential steps: 
\begin{enumerate}
 \item Generation of synthetic git environments
 \item Generation of evaluation requests for each environment
 \item Construction of the ground truth state for each request
 \item Evaluation of the assistant against the ground truth states
\end{enumerate}

The generation of synthetic environments follows two alternatives:

\noindent \textbf{1.a Base Environment}:
This consists of a pair of hand-crafted local and remote repositories, designed to encompass all relevant features necessary for diverse testing. The environment includes:

\begin{itemize}
\item  \textit{Paired branches}: Branches present in both local and remote repositories. The main branch is synchronized, while three additional branches represent cases where local branches are behind, ahead of, or have diverged from their remote counterparts.
\item \textit{Special branches}: One branch that only exists in the remote repository, and another branch that only exists in the local repository.
\end{itemize}

Text files are included in the commits of these branches. Additionally, untracked files and modifications to existing files are introduced in the working directory, which prevents direct checkout on three branches. Functionality test templates are prepared to cover various scenarios (e.g., merging "ahead" into "behind," rebasing "only-remote" onto "in-sync"). Functionalities are grouped into categories such as updating/moving, committing, merging, rebasing, squashing, reverting, and tagging. These templates may also include complementary intents for pulling and pushing changes.

\noindent \textbf{1.b Random Environments}:
These environments are generated by a stochastic state machine that simulates collaborative git workflows, producing random local and remote repositories. It  emulates scenarios where multiple developers interact with both local and remote branches, often resulting in divergent or outdated histories. The system alternates between two perspectives: "mine" (the user’s local copy) and "others" (other developers’ copies). Over 20–30 iterations, the algorithm randomly switches perspectives and performs various git operations according to configurable probabilities (e.g., branch creation p=0.2, commit p=0.8, merge p=0.3, rebase p=0.1, pull p=0.8, push p=0.8, stash p=0.1, dirty index p=0.1).
This probabilistic approach ensures the generation of diverse repository states and branch structures, reflecting typical workflows where local branches may lag behind remote changes or contain uncommitted modifications.


    

Then, we proceed to generate the requests for evaluation.
For the base environment case, we use functionality templates to create 25 requests that cover a range of typical repository management tasks. Then, for each request, we generate four additional versions with the assistance of a language model. Specifically, we use GPT-4o with a tailored prompt to paraphrase the user's request while preserving its original intent. This process yields a total of 100 requests to be addressed which were manually checked, facilitating the subsequent construction of the ground truth. 

For the random environment case, our aim is to evaluate more complex tasks or those that combine multiple functionalities. Here, we manually generate 25 cases (e.g., retrieving updated copies of branches not present in the local repository). As in the previous group, we then produce four equivalent versions for each case using the language model and manually verify them, resulting in a total of 100 requests.

To construct the ground truth data, we manually curate git command sequence that solve each of the requests. For every request, we apply each sequence to either the base or random environment, resulting in a reference state that serves as the ground truth. Notably, this final environment state is shared among all paraphrased versions of a given request, ensuring consistency in evaluation.

For the evaluation, each request is submitted to \Gitassist, which operates on the corresponding environment copy and transitions it to a new final state. The evaluation focuses on comparing the resulting environment states rather than the specific command sequences executed. The comparison encompasses:

\begin{enumerate}
    \item verifying file hashes of local branches and remote branches
    \item verifying file hashes of the working copy
    \item checking the existence of tags in each branch
\end{enumerate}

Intermediate effects on the path to the target state are not considered. For example, committing two files in separate commits is treated as equivalent to combining them in a single commit. Additionally, the stash stack is not considered, as its resulting state (i.e., applying or popping existing stashes) may depend on general user preferences that are not typically specified in individual requests. Nevertheless, by analyzing final state elements, we can provide several levels of matching metrics, which are discussed in the next section.
 
\section{Evaluation}
In this section we evaluate the effectiveness of the two assistant versions, compared to a direct LLM request, named LLM-Vanilla, that we set as our benchmark. The LLM-Vanilla is equal to the LLM-based Engine, but without including the repository state.
Since the human intervention for conflict resolution is not practical
in an automated evaluation, we turn off the interactive mode in all configurations.  In this no-interaction mode, conflicts are resolved preferring the newest commit from the history.  For instance, the git merge command has the option to provide this strategy ($\mathsf{-X [ours\mid theirs]}$).

\begin{table*}[hbt]
\centering
\begin{tabular}{lccccccc}
\toprule 
\textbf{Assistant} & \textbf{Accuracy} & \textbf{Plan Time} & \textbf{Errors} & \textbf{Remote} & \textbf{Local} & \textbf{Working-tree} \\
   & (\%) & (secs) & (\%) & (\%) & (\%) & (\%) \\
\midrule 
LLM-Vanilla & 12.0  & {\bf 3.9} & 97.0 & 37.0 & 29.0 & 17.0 \\

LLM-based & 19.0 & 16.2 & 78.0 & 46.0 & 53.0 & 24.0 \\

Hybrid-Planner & {\bf 81.0} & 31.3 & {\bf 3.0} & {\bf 86.0} & {\bf 85.0} & {\bf 86.0} \\
\bottomrule
\end{tabular}
\caption{Experiment results summary on the base environment dataset}
\label{tab:exp_results_base}
\end{table*}

For this evaluation, we used GPT-4o~\cite{hurst_gpt4o} for all interactions with the LLM.  We carried out additional experiments using the o4-mini model \cite{openai_o3_o4_mini}, that we report on Appendix~\ref{exp:o4mini}. 
To run the planning component of \Gitpddlassist, we used the Fast-Downward Planner~\cite{helmert2006fast}, configured with A* search and the LM-cut heuristic, as provided in the Unified Planning Framework (UPF) Python package~\cite{unified_planning_softwarex2025}. This configuration generates plans that are guaranteed to be valid and optimal. We set a time limit of 60 seconds for the planner. Although some instances may require additional time to be solved, we aim to enable a meaningful comparison with an alternative pure LLM-based solution, which can be generated in a few seconds.

For each configuration, we assess the following metrics:

\begin{itemize}
    \item Accuracy:
The percentage of cases where the executed plan reaches a final state that fully matches the ground truth.
    \item Total Time:
Total time refers to the duration required to generate a git plan, depending on the technique used. It includes the sensing and reasoning capabilities.
    \item Errors: The percentage of tasks in which any error was registered, whether due to git command failure or other execution issues.
    \item State Matches: This metric evaluates the goal state separately as follows: (a) Remote: Whether remote branches and their file hashes coincide with the expected state. (b) Local: Whether local branches and file hashes match with the expected state (c) Working Tree: Whether the working tree matches the expected state. 
\end{itemize}

Table~\ref{tab:exp_results_base} presents the results for the Base environments. The \Gitpddlassist clearly outperforms the two other configurations with a 81\% of accuracy.  Notably, it also produced
the fewest number of errors (3\%), all attributed to the planner not finding a plan for the given goals generated by the LLM.  Regarding the LLM-based, we observe how
including the repository state in the context provides some improvements, more remarkably in the increase of the local branches state matching, as can be seen in the difference of performance between LLM-Vanilla and LLM-based. In LLM-Vanilla, the vast majority of errors (around 92\%) come from failing to checkout to branches without recognizing that git will prevent to do that due to a conflicting working tree with the target branch or pulling without committing or stashing current changes. In LLM-based, the pull errors are mostly solved as the LLM has context of the current state of the working tree, but still most of the errors (around 90\%) come from failing to checkout.
The \Gitpddlassist handle both of these effectively including (git stash) in the plan before trying to switch branches on a dirty working tree.

\begin{table*}[hbt]
\centering
\begin{tabular}{lcccccc}
\toprule
\textbf{Assistant} & \textbf{Accuracy} & \textbf{Plan Time} & \textbf{Errors} & \textbf{Remote} & \textbf{Local} & \textbf{Working-tree} \\
   & (\%) & (secs) & (\%) & (\%) & (\%) & (\%) \\
\midrule
LLM-Vanilla & 12.0 & {\bf 5.5} & 89.0 & 57.0 & 48.0 & 21.0 \\

LLM-based & 16.0 & 20.2 & 76.0 & 53.0 & 41.0 & 25.0 \\

Hybrid-planner & {\bf 59.0} & 38.4 & {\bf 10.0} & {\bf 81.0} & {\bf 74.0} & {\bf 60.0} \\
\bottomrule
\end{tabular}
\caption{Experiment results summary on the random environment dataset}
\label{tab:exp_results_random}
\end{table*}

Regarding response time, LLM-Vanilla is the fastest, as its input prompt is smaller due to not including the repository state in the context. The LLM-based engine has an average response time of 16.2 seconds.The Hybrid-planner incurs an additional overhead of approximately 15 seconds compared to the LLM-based engine, primarily due to the time required by the solver to generate the plan. Nevertheless, the total time remains under one minute, which is considered manageable. Developers will still benefit, as after confirmation, the bulk of commands are executed in an average of 10.5 seconds. We also analyzed performance across different functionalities or operation classes. Results are shown in Table~\ref{tab:result_per_class}. Considering the distribution of tasks,  for classes where the hybrid-planner did not achieve perfect accuracy, there was a single failure per class. These errors were due to the LLM’s interpretation of one rephrased request, while the other three were handled correctly.

\begin{table}[]
    \centering
    \footnotesize{
 \begin{tabular}{lrrrr}
\toprule
Assistant & Num & Vanilla & LLM & Hybrid \\
operation &  &  &based  &planner \\
\midrule
branch creation & 12  & 8.3 & 33.3 & 100.0 \\
commit & 16 & 0 & 50.0 & 100.0 \\
merge & 16 & 0.0 & 0.0 & 100.0  \\
rebase & 20 & 10.0 & 15.0 & 95.0 \\
reset & 8 & 0.0 & 37.5 & 87.5 \\
squash & 8 & 0.0 & 0.0 & 100.0 \\
tagging & 12 & 0.0 &  33.3 & 91.7 \\
update and move & 8& 0.0 & 0.0 & 100.0 \\
\bottomrule
\end{tabular}}
    
    \caption{Accuracy per git operation class obtained in the base environment dataset}
    \label{tab:result_per_class}
\end{table}

We also evaluated the performance on the random environment dataset. Table~\ref{tab:exp_results_random} shows the results for this experiment. In this case, the \Gitpddlassist again obtained the best performance with a 59\% of accuracy. The rate is lower as the type of tasks are more complex than in the base environment.
In these results we see that the comparison of the working tree state tends to be the limiting factor in terms of overall accuracy. Analyzing the errors, we observe that the Hybrid-planner may still fail due to the inability to find a valid plan for the set of goals generated by the LLM. In contrast, the LLM-based and LLM-Vanilla approaches predominantly encounter previously mentioned errors, such as attempting to perform a pull, checkout, or merge in a modified (dirty) working tree. Both approaches also exhibit errors related to hallucinating command options that do not exist (e.g., using “--no-edit” during a rebase).

Since the working tree metric generally yields the poorest results, it could be inferred that its final state is often obscured in the user's request. Under this assumption, we may focus on the remote metric, as developers explicitly specify what they intend to share with others. Nevertheless, in this case, the conclusion is still similar: LLM-Vanilla and LLM-based approaches show comparable performance, while the Hybrid-planner achieves superior results that exceed both by at least 30 percentage points.

\section{Discussion and Limitations}
Although \Gitpddlassist covers the typical functionalities of git, the universe of Git commands and parameters extends far beyond, enabling the handling of much more complex scenarios. Among the functionalities we have excluded, it is important to highlight the following:

\begin{itemize}
\item Cloning and Configuration: We assume the repository already exists, as these operations are infrequent and provide limited value to the assistant.
\item File Restoration: The git restore command, which allows recovery of a file’s version from a previous commit, is typically performed individually per file and is often accessible with a single click in IDE plugins. It does not require to be handled as part of a multi-command plan.
\item Difference Analysis: The git diff command does not alter the repository state, and its potential is best exploited through visual comparison in the IDE.
\end{itemize}

Furthermore, we have not addressed the management of deleted files across branches, unlike our treatment of modified and untracked files. This functionality is reserved for future work. Another relevant point is that the assistant assumes atomic operations; however, developers sometimes seek assistance to recover from unstable states, such as continuing an interrupted or corrupted merge. Representing such scenarios as initial states requires more detailed analysis.

The advantage of the \Gitpddlassist lies in its reliability, as none of its failures comes from incorrect git command execution. This allows developers to trust that the proposed commands are safe to apply and that the resulting states will be stable throughout the process.

Finally, the reported times correspond to the git environments used in the evaluation. Larger repositories imply that the Observer will require more time, as most information is extracted using native git commands whose performance scales with repository size. We expect the planning time to be less affected, since most of the relevant data comes from the latest commit and branches, while typical projects tend to grow more rapidly in terms of files and commits.

\section{Conclusion}
This work introduces \Gitassist, a hybrid AI tool that integrates LLMs with automated planning to support developers in executing complex git operations. Through systematic evaluation in both synthetic and randomized environments, our results demonstrate that combining formal reasoning with LLMs significantly improves reliability and reduces errors in repository management tasks. The hybrid planning approach, in particular, offers robust handling of non-trivial scenarios and enhances user trust by ensuring safe and correct command sequences. While some advanced git functionalities remain outside the current scope, our findings highlight the potential of hybrid AI assistants to lower the barrier to effective version control.

\section*{Disclaimer}
This paper was prepared for informational purposes by the Artificial Intelligence Research group of JPMorgan Chase \& Co. and its affiliates ("JP Morgan'') and is not a product of the Research Department of JP Morgan. JP Morgan makes no representation and warranty whatsoever and disclaims all liability, for the completeness, accuracy or reliability of the information contained herein. This document is not intended as investment research or investment advice, or a recommendation, offer or solicitation for the purchase or sale of any security, financial instrument, financial product or service, or to be used in any way for evaluating the merits of participating in any transaction, and shall not constitute a solicitation under any jurisdiction or to any person, if such solicitation under such jurisdiction or to such person would be unlawful.

© 2026 JPMorgan Chase \& Co. All rights reserved

\bibliographystyle{named}
\bibliography{references}

\clearpage

\appendix
\section{Appendix}

\subsection{Prompts}
\label{appendix:prompts}

\subsubsection{LLM-based Engine Template Prompt ($T_1$)}
\begin{tcolorbox}
\begin{small}
You are an expert git assistant. The user has requested:

**USER INPUT **

Your task is to generate a sequence of git commands to fulfill the user's objective.

INSTRUCTIONS:

1. Analyze the current repository state

2. Plan the necessary command sequence

3. Show the sequence in python format (a list of commands)

4. Briefly explain the purpose of each command

NOTES:

- Do not include comments in the python code, just the commands to be executed

- local files as groups by untracked, modified, staged, etc. for adding/commiting

- only add modified files unless the request specifies to add 'all' or 'untracked' files

- when pushing or pulling, assume the remote is named 'origin'

- Everytime the request refers to branches that are not the one you are currently in, syncronize with remote (git fetch + merge or pull) those banches unless otherwise stated. This mean those branches are from other developer (in remote) and are required to be sync with latests changes prior of using them.
For the one you are currently in, it depends on the action to be done. For example, if specified "current" or "local", you should not syncronize. Only syncronize the required branches. The ones that are not used or not required, keep it state as it is.

- Before doing a push or similar, think always if a prior pull or similar is required to avoid errors.

- All commands are going to be executed from the repo root path.

** REPOSITORY CONTEXT **

** LAST COMMITS INFO **

** GIT GRAPH INFO **

** INTERACTIVE INSTRUCTIONS **

** FEW SHOT EXAMPLES **

Return only the code block with the list of commands and the explanation.
\end{small}
\end{tcolorbox}

\newpage
\subsubsection{Hybrid-Planner Instruction Template Prompt ($T_2$)}
\begin{tcolorbox}
\begin{small}

You are a git assistant. But instead of you providing a sequence of git commands,
I will use an external PDDL planner to create the plan that contains the actions
to execute. Your task is to understand the user request and  match the intent
to the PDDL goals that will achieve them.

**PPDL PREDICATE DEF REFERENCE **

TASK:
Your task is to produce a python snippet containing a dictionary containing:

goal: a list of strings with single pddl atoms representing the needed goals

mapping: another dictionary to match pddl objects to real items,
such as branch names, specific files, tag names, etc. 

To help reasoning about the goals you need
to write first your interpretation of the user intent.

NOTES:

- The interpretation is a list of each recognizable sub-goals

- The goals must be achievable using only git commands

- The mapping is only needed when the request refers to tag names, new branches, or specific file names

- When including goals remember to follow the order of types in the PDDL predicate. For example, in (sync ?lb - local\_branch ?rb - remote), the relevant local\_branch comes first and then the corresponding remote branch.

 - If you explicitly recognize that the request cannot be done or it is a bad practice, respond with an empty goal list and an empty mapping. However, it's preferred to handle this with the planner, so the problem would be unsolvable as a proof.

- If there are ambiguities in the request regarding local or origin branches, prefer the that  branches are in sync and updated.

 - for operations with local branch X, when it is behind or has diverged, include updated goals (update X) unless instructed not to do it. However, a request for fetching from origin does not mean to update all branches, only the ones required for the task.
 
 - When branches are sync or ahead with origin, they have the latest changes, so no need to include updated goals for them.

 - if the working copy has modified files and changes are not mentioned to be committed or merged, include a goal to keep the working copy (modified) state to prevent forgetting intermediate stashes. Do not include working copy goal if something is committed in the task.  Do not include this goal is the working copy is clean.
 
 **FUNCTIONALITY INSTRUCTIONS**

**FORMAT INSTRUCTIONS **

Suppose for the example these facts are in the initial state

**INITIAL STATE EXAMPLE**

Now I provide examples of user request in the form

INPUT: user request

OUTPUT: interpretation + python snippet 

** FEW SHOT EXAMPLES**

Now from this point this is REAL DATA. The PDDL representation of the git repo as a PDDL partial problem is

** INIT STATE IN PDDL**

THE USER REQUEST IS: **REQUEST**

REMEMBER TO ONLY OUTPUT THE INTERPRETATION AND PYTHON SNIPPET AS IN THE EXAMPLES
\end{small}
\end{tcolorbox}

Additionally, the following snippet show an example included in the **FEW EXAMPLE** placeholder above.
\begin{tcolorbox}
    \begin{small}
INPUT:

"Add 'fileA.txt' and 'fileB.txt', commit and push to origin"

OUTPUT:

Interpretation:

    - commit files (fileA.txt, fileB.txt) as a specific content
    
    - last commit on current branch feat1 is pushed to origin/feat1
    
```python

\{

\hspace*{1em} "goal": [
    
\hspace*{2em} "(commited\_content c\_new0 feat1 specific)",
        
\hspace*{2em} "(last\_commit origin/feat1 c\_new0)"
        
\hspace*{1em} ],
    
\hspace*{1em} "mapping": \{
    
\hspace*{2em} "specific": ["fileA.txt", "fileB.txt"]
        
 \hspace*{1em}\}
    
\}

```
    
    \end{small}
\end{tcolorbox}

\subsection{Git Domain}
\label{appendix:git-domain}
The PDDL Git Domain represents the repository mechanics required to track git operations in local and remote environments. Each action listed in Table~\ref{tab:git-tasks-groups} includes: (1) a list of parameters (objects involved in the operation), (2) preconditions (facts that must be true in the state for the action to be applicable), and (3) effects (the changes in the state after the action is applied). Below is an example of a complete PDDL action.

\begin{tcolorbox}
\footnotesize{
(:action \textbf{create\_branch}

\hspace*{1em}    :parameters (?newb - local\_branch ?b - local\_branch

\hspace*{6em} \ ?c - commit)

\hspace*{1em}    :precondition (and (current\_branch ?b)

\hspace*{7em} (last\_commit ?b ?c) 

\hspace*{7em} (not (in\_repo ?newb))
                       
\hspace*{7em} (unused\_branch ?newb))

\hspace*{1em}:effect (and (in\_repo ?newb)

\hspace*{4em}    (last\_commit ?newb ?c)
                 
\hspace*{4em}    (created\_from ?newb ?b)
                 
\hspace*{4em}    (current\_branch ?newb) (not 

\hspace*{4em}    (current\_branch ?b)))

)
}
\end{tcolorbox}

\subsection{Reasoning model experimentation}
\label{exp:o4mini}
We have also carried out the experiments on bot base and random environment using a so called "reasoning" LLM, in this scenario the model "o4-mini" from OpenAI~\cite{openai_o3_o4_mini} which is comparable to the GPT-4o used in terms or released date a capabilities. Taking a look at the results in Tables~\ref{tab:exp_results_base_reason} and \ref{tab:exp_results_random_reason}, we can observe how in this scenario the LLM-Vanilla and LLM-based results tend to have a better accuracy compared to the results using GPT-4o. In contrast, the accuracy of for the Hybrid-Planner gets a significant decrease in accuracy but still it beats the other two assistants. Overall, the plan time suffers an increase in all scenarios due to the nature of the reason models that need more time to do so. Moreover, the distribution of types of errors is very similar to the error analysis done with GPT-4o

Trying to extract some insights, we can hypothesize that these results are related to the nature of the task to be carried out and the nature of the models. The o4-mini is a "reasoning" model that tends to do better in tasks that require some thinking and steps, like maths, code, planning and so on. On the other hand, the GPT-4o is a base model that tends to do better in general tasks like semantic understanding, knowledge, chat and so on.

In our experimentation, the task to be carried out in the assistants LLM-Vanilla and LLM-based is related to the reasoning and planning field where the LLM has to develop a sequential plan formed by git commands to satisfy the user request. Here we observe that the "reasoning" model o4-mini tends to obtain better results. On the other hand, the task to be carried out in the assistant Hybrid-planner is related to semantic understanding, where the LLM has to understand the user intention and represent it as goals. Here, we observe that the base model GPT-4o excels, surpassing by a significant margin its counterpart using o4-mini. 
Many of the mismatches arise from failing to match the working tree with the ground truth. In these cases, no execution errors are generated, but the o4-mini takes liberties in interpreting how to leave the working tree, especially regarding the handling of stashes.

\begin{table*}[hbt]
\centering
\begin{tabular}{lccccccc}
\toprule 
\textbf{Assistant} & \textbf{Accuracy} & \textbf{Plan Time} & \textbf{Errors} & \textbf{Remote} & \textbf{Local} & \textbf{Working-tree} \\
   & (\%) & (secs) & (\%) & (\%) & (\%) & (\%) \\
\midrule 
LLM-Vanilla & 19.0 & 12.9 & 91.0 & 50.0 & 48.0 & 20.0 \\

LLM-based & 24.0 & 31.9 & 65.0 & 62.0 & 58.0 & 26.0 \\

Hybrid-Planner & 34.0 & 44.5 & 1.0 & 76.0 & 71.0 & 46.0 \\
\bottomrule
\end{tabular}
\caption{Experiment results summary on the base environment dataset using o4-mini}
\label{tab:exp_results_base_reason}
\end{table*}

\begin{table*}[hbt]
\centering
\begin{tabular}{lcccccc}
\toprule
\textbf{Assistant} & \textbf{Accuracy} & \textbf{Plan Time} & \textbf{Errors} & \textbf{Remote} & \textbf{Local} & \textbf{Working-tree} \\
   & (\%) & (secs) & (\%) & (\%) & (\%) & (\%) \\
\midrule
LLM-Vanilla & 9.0 & 16.1 & 88.0 & 59.0 & 41.0 & 21.0 \\

LLM-based & 26.0 & 32.8 & 50.0 & 68.0 & 49.0 & 35.0 \\

Hybrid-planner & 42.0 & 48.2 & 17.0 & 79.0 & 71.0 & 44.0 \\
\bottomrule
\end{tabular}
\caption{Experiment results summary on the random environment dataset using o4-mini}
\label{tab:exp_results_random_reason}
\end{table*}

\end{document}